\documentclass[twocolumn,aps,prb,longbibliography]{revtex4-2}
\usepackage{amsmath}
\usepackage{amssymb}
\usepackage{amsfonts}
\usepackage{bm}
\usepackage{physics}
\usepackage{graphicx}
\usepackage{grffile}
\usepackage{float}
\usepackage{placeins}
\usepackage{subcaption}
\usepackage{xcolor}
\usepackage{booktabs}
\usepackage[colorlinks=true,linkcolor=blue,citecolor=blue,urlcolor=blue]{hyperref}

\begin{document}

\title{Helical Quasiperiodic Chains with Engineered Dissipation: Liouvillian Rapidity Diagnostics of Transport and Localization}

\author{Mohammad Pouranvari}
\affiliation{Department of Physics, Faculty of Basic Sciences, University of Mazandaran, Babolsar, Iran}

\date{\today}

\begin{abstract}
  We study relaxation spectra of a quadratic spinless-fermion ``helical'' chain with an Aubry–André–type quasiperiodic potential and a single $N$-th neighbor (helical) hopping. Dissipation and pumping are introduced via local linear Lindblad jump operators and treated exactly using the third-quantization / Majorana covariance formalism. Focusing on periodic boundary conditions (to avoid edge artefacts) we compute the Liouvillian rapidities and their smallest nonzero real part (the rapidity gap) for several spatial dissipation patterns: uniform (`all'), single-site (`one-site') and two-site (`two-site') placement, plus pairwise gain/loss on helical partner sites. We show that uniform dissipation yields large, weakly $\lambda$-dependent gaps, while sparse local dissipation produces gaps that shrink rapidly as the quasiperiodic potential $\lambda$ induces localization. Increasing $t_N$ enhances relaxation by improving mode overlap with dissipative channels. Finite-size scaling, rapidity level statistics (Poisson vs Wigner–Dyson), and spatial profiles of slow modes provide a consistent picture linking Liouvillian spectral structure to transport and localization. Our results highlight Liouvillian rapidities as compact, experimentally relevant diagnostics of relaxation and sensitivity in engineered open quantum lattices.
\end{abstract}

\maketitle

\section{Introduction}
Open quantum systems governed by Markovian master equations appear across quantum optics, cold atoms, and condensed-matter platforms. The Gorini–Kossakowski–Sudarshan–Lindblad structure provides a general and widely used generator for Markovian dynamics \cite{Gorini1976,Lindblad1976}. When the system Hamiltonian is quadratic and the jump operators are linear in fermionic operators, the steady state and relaxation spectrum can be obtained exactly via a `third-quantization' approach mapping the Liouvillian to a linear problem in Majorana covariance space \cite{Prosen2008,Prosen2010,Seligman2010}. This powerful machinery makes quadratic open systems ideal testbeds to study how dissipation and driving shape nonequilibrium phases, prepare nontrivial steady states \cite{Diehl2008,Kraus2008,Verstraete2009,Poyatos1996}, and generate spectrally sharp features such as Liouvillian exceptional points \cite{Wiersig2014,ElGanainy2018,Chen2017}.

Quasiperiodic lattices — Aubry–André type models in one dimension — combine analytical tractability with localization physics distinct from random disorder: a single-parameter tuning can move the system from extended to localized regimes without ensemble averaging \cite{Aubry1980,Roati2008,Lahini2009,Biddle2010}. Helical long-range hopping (a single $N$-th neighbor term connecting sites $j$ and $j+N$) introduces an additional length-scale and interference pathways that can modify band structure and localization properties. Combining quasiperiodicity with engineered dissipation allows us to ask how localization, long-range hopping, and spatial structure of dissipators jointly determine relaxation time scales and steady-state sensitivity — questions of both fundamental and experimental interest \cite{Diehl2008,Verstraete2009,Poyatos1996}.

This work focuses on Liouvillian rapidities (eigenvalues of the effective damping matrix in the Majorana representation) and in particular the smallest nonzero real part, $\kappa$, which sets the slowest decay rate of single-particle correlations (the Liouvillian gap). Rapidities have emerged as natural diagnostics of relaxation dynamics and sensitivity in open quadratic systems \cite{Prosen2008,Prosen2010,Znidaric2015,Mori2020}. We use them here to characterize regimes of fast vs.\ slow relaxation, level-statistics crossovers (Wigner–Dyson to Poisson) associated with delocalization vs localization \cite{Mehta1991,Haake2010}, and slow-mode spatial structure (mode weight concentrated on dissipative sites implies long-lived modes).

We organize the paper as follows. In Sec.~\ref{sec:model} we introduce the helical quasiperiodic Hamiltonian and the classes of linear Lindblad jump operators we study. Section~\ref{sec:method} summarizes the Majorana / third-quantization mapping and the numerical protocol for rapidity extraction, finite-size scaling, level-statistics and mode characterization. Section~\ref{sec:results} presents the numerical results (rapidity linecuts vs $\lambda$ for several $t_N$, finite-size scaling, spacing statistics, and mode profiles) and physical interpretation. We conclude in Sec.~\ref{sec:conclusions} with implications and possible experimental tests. Related theoretical and experimental background motivating the methods and observables used here can be found in Refs.~\cite{Aubry1980,Roati2008,Lahini2009,Biddle2010,Prosen2008,Diehl2008,Verstraete2009,ElGanainy2018}.

\section{Model}
\label{sec:model}

We consider spinless fermions on a one-dimensional chain of $L$ sites with nearest-neighbour hopping $t$, a single $N$-th neighbour (helical) hopping of amplitude $t_N$, and a quasiperiodic onsite potential of amplitude $2\lambda$. Working in second-quantised form (periodic boundary conditions throughout), the single-particle Hamiltonian reads
\begin{align}
  H &= \sum_{j=1}^L V_j c_j^\dagger c_j - t\sum_{j=1}^L (c_j^\dagger c_{j+1} + \mathrm{h.c.}) \nonumber\\
  &\quad - t_N\sum_{j=1}^L (c_j^\dagger c_{j+N} + \mathrm{h.c.}),
  \label{eq:H}
\end{align}
with $c_{L+1}=c_1$ (PBC). The onsite potential is quasiperiodic,
\begin{equation}
  V_j = 2\lambda\cos(2\pi b j + \phi),
\end{equation}
where $b$ is chosen incommensurate (we use the inverse golden ratio $b=(\sqrt{5}-1)/2$ in numerics) and $\phi$ is a uniform random phase used to average sample-to-sample fluctuations where appropriate \cite{Aubry1980,Roati2008}.

Dissipation and pumping are included via a Lindblad master equation
\begin{equation}
  \dot\rho = -i[H,\rho] + \sum_m\Big(L_m\rho L_m^\dagger - \tfrac{1}{2}\{L_m^\dagger L_m,\rho\}\Big),
\end{equation}
and we restrict to \emph{linear} jump operators of the local form
\begin{equation}
  L_j = \sqrt{\gamma_j}\,(\alpha_j c_j + \beta_j c_j^\dagger),
  \label{eq:Lj}
\end{equation}
with complex coefficients $\alpha_j,\beta_j$ (allowing both loss and gain, phases, and linear combinations). This class preserves Gaussianity and admits exact third-quantization treatment \cite{Prosen2008,Kraus2008,Diehl2008,Verstraete2009}.

We analyze several spatial placement patterns for the $L_j$: uniform on all sites (`all'), a single dissipator on one site (`one-site'), two dissipators on two distinct sites (`two-site'), alternating gain/loss across the lattice, and pairwise gain/loss at helical partners $(j,j+N)$. These placements model experimentally relevant reservoir engineering patterns (e.g., spatially patterned pumping/absorption) \cite{Poyatos1996,Kraus2008}.

\section{Method}
\label{sec:method}

\subsection{Majorana representation and the damping matrix}
Introduce Majorana operators $w_{2j-1}=c_j + c_j^\dagger,\; w_{2j}=i(c_j - c_j^\dagger)$ and collect them into a real $2L$ vector $w$. The Hamiltonian is represented as $H=\tfrac{i}{4}w^T H_M w$ with $H_M$ real and antisymmetric. Each linear jump operator can be written $L_m=\sum_p \ell_{m,p} w_p$ and one defines the bath matrix
\begin{equation}
  M_{pq}=\sum_m \ell_{m,p}\ell_{m,q}^*,
\end{equation}
which is Hermitian and positive semidefinite. Following third-quantization \cite{Prosen2008,Prosen2010,Seligman2010} the key non-Hermitian rapidity matrix is
\begin{equation}
  X = -4i H_M + 2(M + M^T),
  \label{eq:X}
\end{equation}
and one also defines $Y=-4i(M-M^T)$ which enters the Lyapunov equation below. The eigenvalues of $X$ are the rapidities $\{\beta_k\}$: their real parts govern decay rates and the smallest nonzero $\Re\beta$ is the Liouvillian rapidity gap $\kappa$ (we report $\kappa$ as the central dynamical observable) \cite{Prosen2008,Znidaric2015}.

\subsection{Steady covariance and Lyapunov equation}
The steady-state Majorana covariance matrix $\Gamma$ satisfies a Sylvester/Lyapunov equation
\begin{equation}
  X\Gamma + \Gamma X^T = -4i(M - M^T) \equiv Y,
  \label{eq:lyap}
\end{equation}
which is solved numerically using standard linear algebra routines for each parameter point to obtain steady correlations and observables such as $\langle c_j^\dagger c_j\rangle$ \cite{Prosen2008,Prosen2010}. The same $X$ controls rapidities and mode structure.

\subsection{Diagnostics and numerics}
For each choice of $N$, $t_N$, and $\lambda$ (and for each dissipator placement pattern) we compute:
\begin{enumerate}
\item The Liouvillian rapidities (eigenvalues of $X$) and extract the smallest nonzero $\kappa=\min_{k\ne 0}\Re\beta_k$.
\item Finite-size scaling of $\kappa(L)$ for $L$ up to a few hundred when feasible, to assess thermodynamic trends \cite{Znidaric2015,Mori2020}.
\item Level-statistics of the rapidities' real parts: nearest-neighbour spacing distribution $P(s)$ (unfolded) compared to Poisson and Wigner–Dyson (GOE) benchmarks; crossovers from WD to Poisson indicate crossover from chaotic/delocalized to localized spectral statistics \cite{Mehta1991,Haake2010}.
\item Spatial weight of the slowest-decaying mode (mode-localization diagnostic): grouping Majorana components per site yields site-resolved weights.
\end{enumerate}
Numerics use dense diagonalization of $2L\times 2L$ matrices for $X$ (possible for the sizes analyzed here) and averaging over a small number of $\phi$ realizations where indicated.

\section{Results}
\label{sec:results}

We present rapidity linecuts $\kappa(\lambda)$ for representative placements and $t_N$ values, finite-size scaling and spectral diagnostics for selected $\lambda$'s (one in the delocalized-like regime and one in the localized-like regime), and slow-mode spatial profiles. All data use periodic boundary conditions.

\subsection{Rapidity linecuts: dependence on $\lambda$ and $t_N$}
\begin{figure*}[t]
  \includegraphics[width=0.32\linewidth]{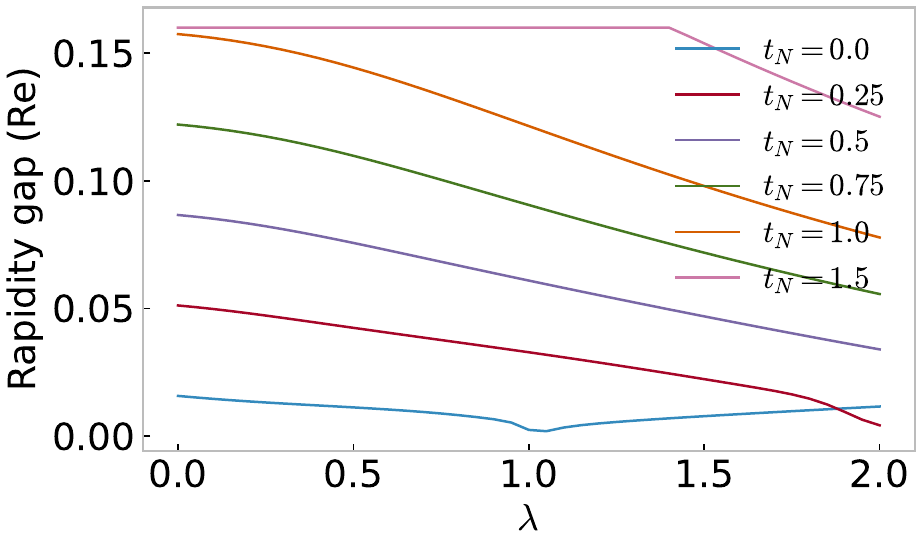}
  \includegraphics[width=0.32\linewidth]{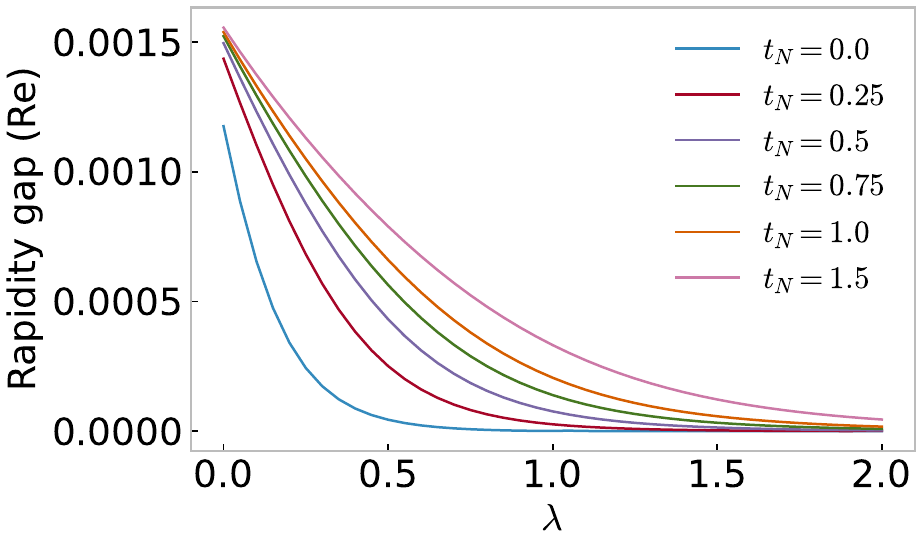}
  \includegraphics[width=0.32\linewidth]{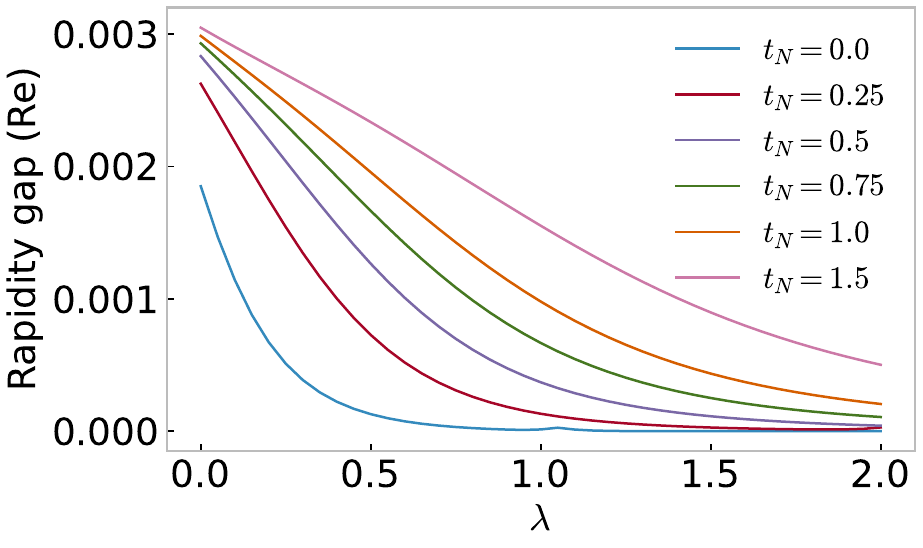}
  
  \caption{Linecuts of the Liouvillian rapidity $\kappa$ versus quasiperiodic potential strength $\lambda$ for several values of the long-range hopping amplitude $t_N$. Each curve corresponds to one $t_N$ (legend in each panel). All data were obtained under periodic boundary conditions. Shown are: (a) uniform dissipation on all sites, (b) single-site dissipation and (c) two-site dissipation. Uniform dissipation yields the largest rapidity gaps and only weak dependence on $\lambda$, reflecting strong coupling of all modes to the environment. In contrast, localized dissipation leads to much smaller gaps that decrease rapidly with increasing $\lambda$, consistent with localization suppressing the overlap between slow modes and dissipative sites. Increasing $t_N$ systematically enhances $\kappa$, indicating that long-range hopping improves transport to dissipators and accelerates relaxation.}
  \label{fig:rapidity_summary}
\end{figure*}

Figure~\ref{fig:rapidity_summary} shows linecuts of $\kappa$ vs $\lambda$ for $t_N\in\{0,0.25,0.5,0.75,1.0,1.5\}$ and the placement patterns that give nonzero gaps in our scans: uniform (`all'), single localized dissipator (`one-site'), and two localized dissipators (`two-site'). The main trends are:
\begin{itemize}
\item \textbf{Uniform (`all')}: large rapidity gaps (order $10^{-1}$ in our units) and only weak dependence on $\lambda$. Dissipative channels cover the lattice so localization does not drastically reduce overlap with dissipators.
\item \textbf{One-site / Two-site}: much smaller gaps that shrink rapidly as $\lambda$ increases. This reflects that localization reduces overlap between the slow modes and the sparse dissipators, producing long relaxation times $\tau\sim 1/\kappa$.
\item \textbf{Effect of $t_N$}: larger $t_N$ consistently increases $\kappa$, indicating that long-range hopping enhances mode delocalization or mode transport to dissipative sites.
\end{itemize}
These findings are robust over a range of system sizes (see finite-size scaling below) and align with the intuitive picture that relaxation is set by the overlap of slow modes with dissipative operators \cite{Prosen2008,Znidaric2015}.

\subsection{Finite-size scaling, level statistics and mode profiles}
To explore spectral structure associated with the linecut trends we compiled compact 2×2 diagnostic figures for representative $\lambda$ values (one in the weak-quasiperiodicity regime, one in the strong-quasiperiodicity regime). Each 2×2 panel shows (i) finite-size scaling $\kappa(L)$ for various $t_N$, (ii) ordered real parts of rapidities (highlighting modes near zero), (iii) level-spacing distribution $P(s)$ for the rapidities' real parts compared to Poisson and GOE (Wigner–Dyson) references, and (iv) slow-mode site weight profile for the slowest-decaying mode.

\begin{figure*}[t]
  \centering
  \includegraphics[width=0.32\linewidth]{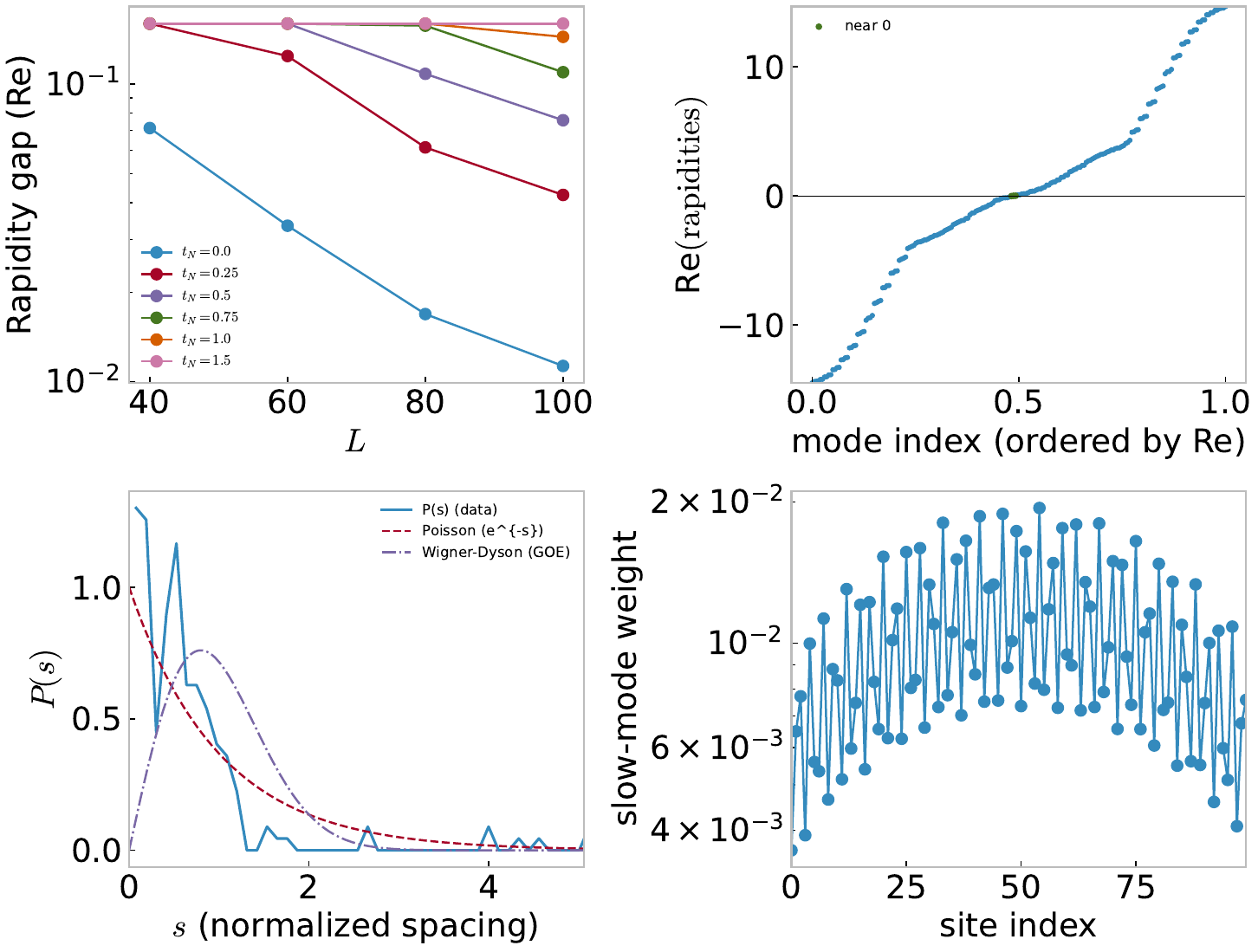}
  \includegraphics[width=0.32\linewidth]{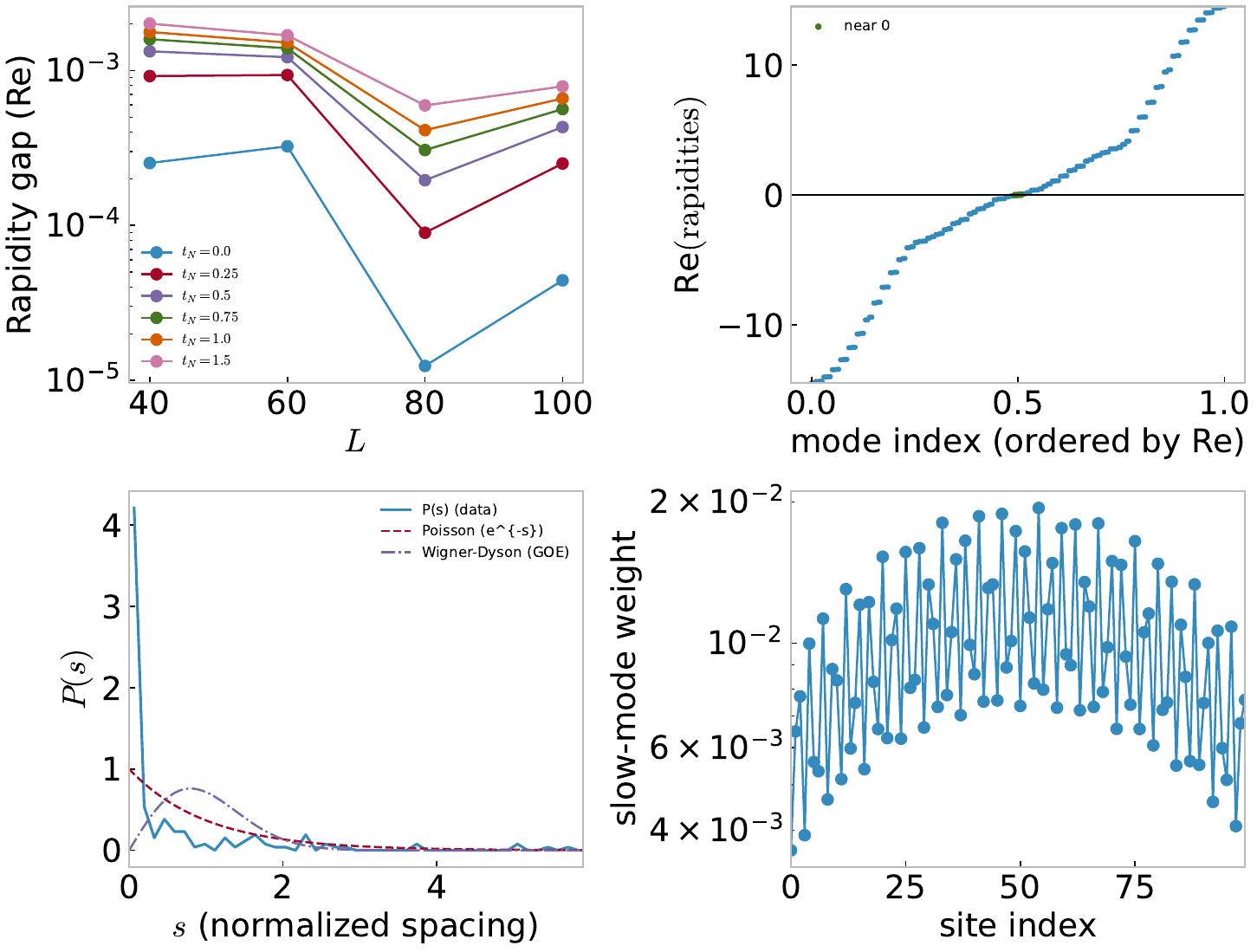} 
  \includegraphics[width=0.32\linewidth]{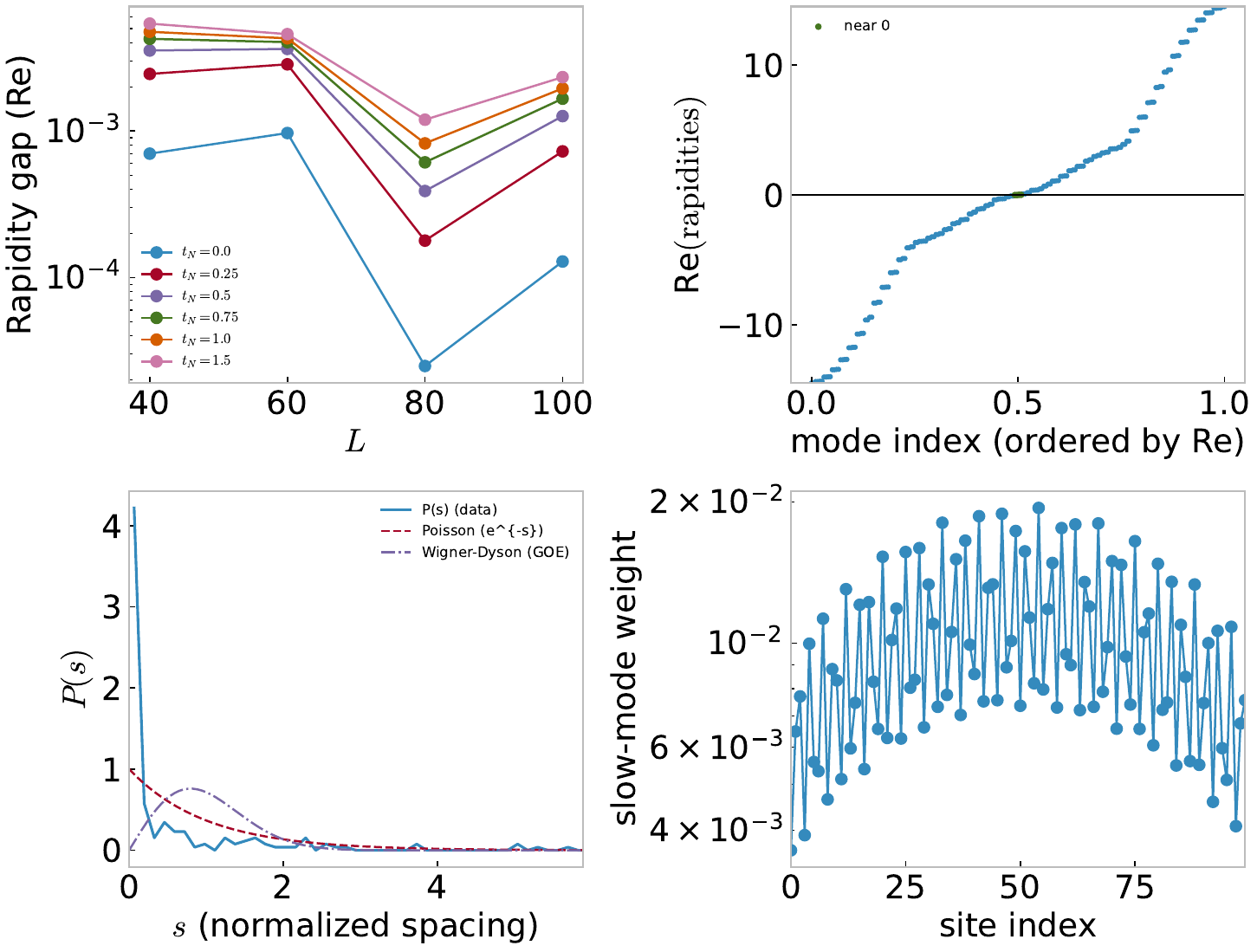}	
  \caption{Spectral diagnostics at $\lambda=0.5$ (transport-dominated regime), $N=3$. Each PDF (panels left to right) is a compact 2×2 diagnostic combining: (top-left) finite-size scaling of the Liouvillian rapidity gap $\kappa$ vs chain length $L$ for several $t_N$ values; (top-right) ordered real parts of rapidities (highlighting modes near zero); (bottom-left) nearest-neighbour spacing distribution $P(s)$ for the rapidities' real parts with Poisson (dashed) and Wigner–Dyson GOE (dot-dash) references; and (bottom-right) site-resolved weight of the slowest-decaying mode. The three panels show, from left to right: (a) uniform dissipation on all sites; (b) a single localized dissipator ; and (c) two localized dissipators. At $\lambda=0.5$ the uniform case displays the largest gaps and spacing statistics closer to Wigner–Dyson (indicative of strong mode mixing), while sparse localized dissipators show smaller gaps and more extended-to-intermediate localization signatures.}
  \label{fig:diagnostics_lambda05}
\end{figure*}

\begin{figure*}[t]
  \centering
  \includegraphics[width=0.32\linewidth]{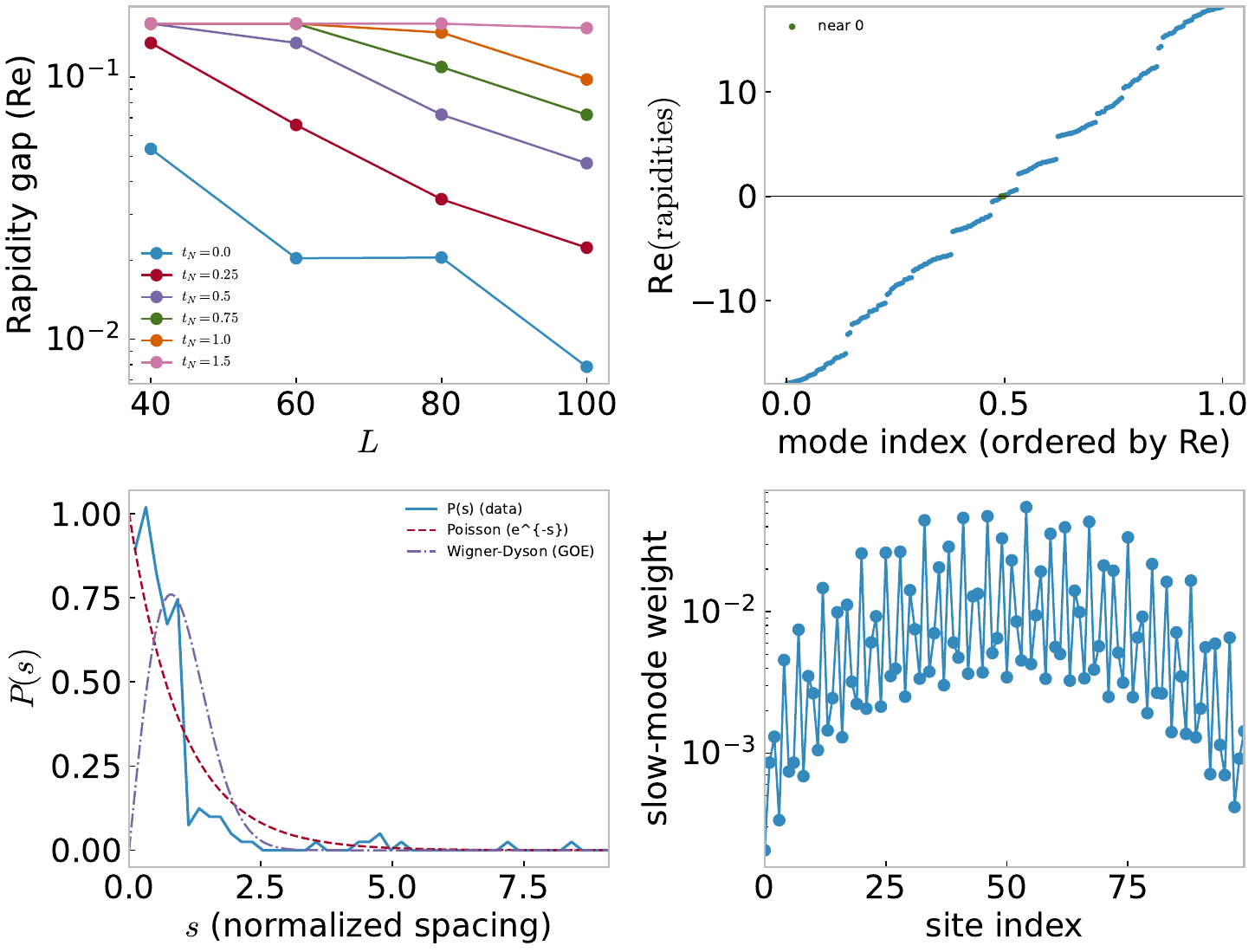}
  \includegraphics[width=0.32\linewidth]{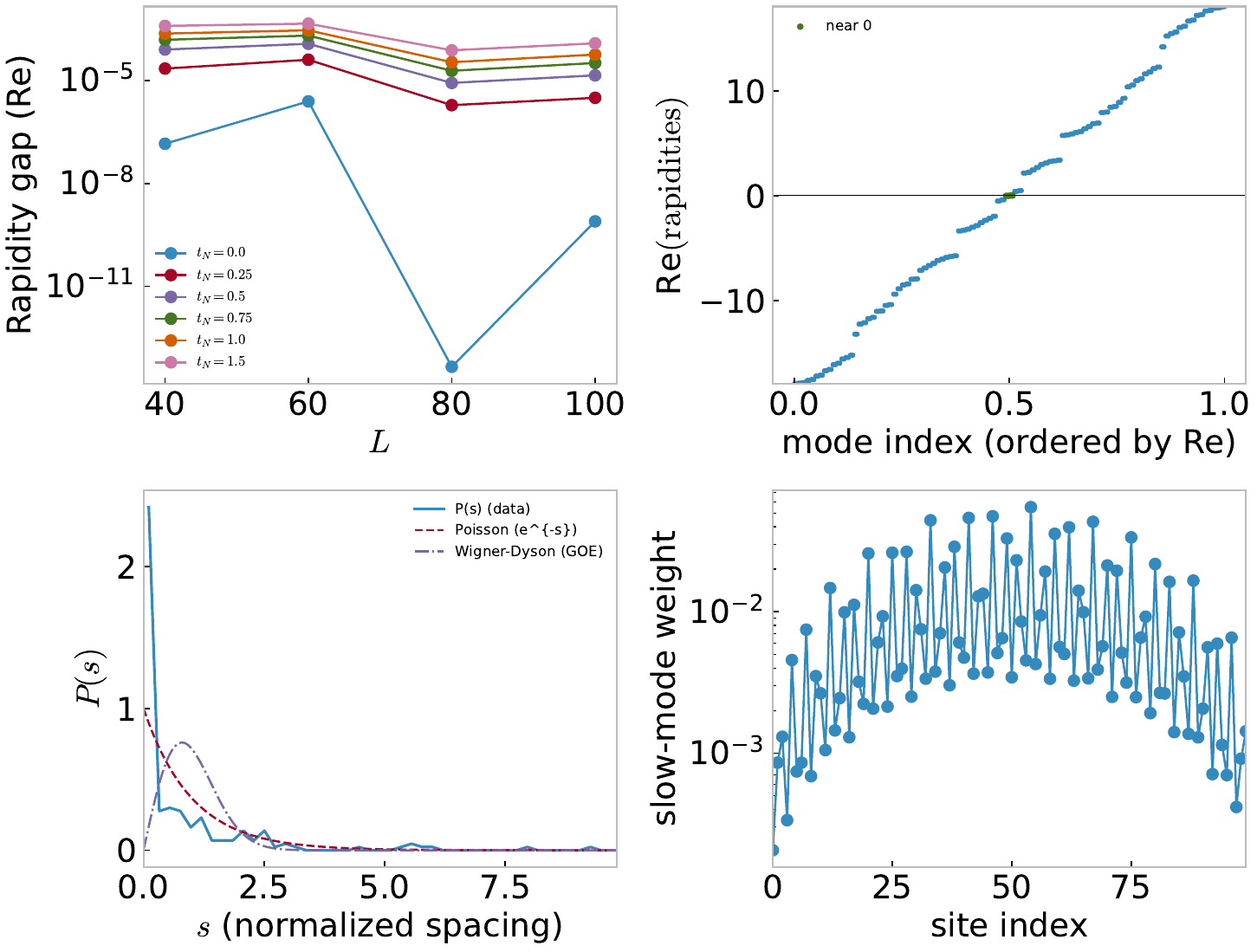} 
  \includegraphics[width=0.32\linewidth]{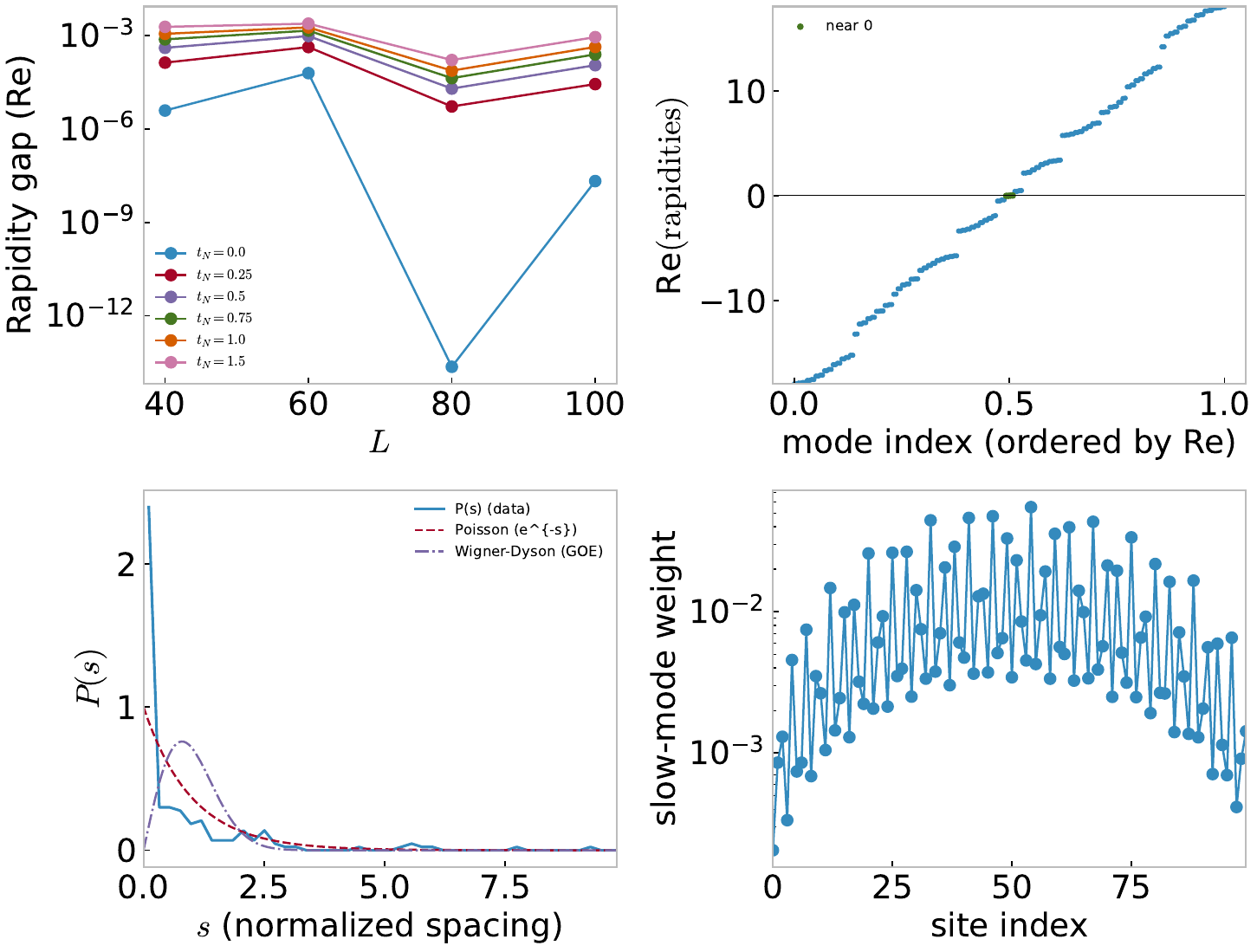} 
  \caption{Spectral diagnostics at $\lambda=1.5$ (localized regime), $N=3$. Panel layout and diagnostic content are the same as in Fig.~\ref{fig:diagnostics_lambda05}. From left to right the panels correspond to: (a) uniform dissipation on all sites; (b) a single localized dissipator ; and (c) two localized dissipators. Increasing $\lambda$ to 1.5 drives stronger localization: sparse dissipator placements exhibit substantially reduced rapidity gaps, spacing distributions closer to Poisson, and slow-mode profiles that concentrate weight on the dissipative site(s). The uniform (`all') placement retains relatively large $\kappa$ and more extended slow modes by comparison.}
  \label{fig:diagnostics_lambda15}
\end{figure*}

At small $\lambda$ we find spacing distributions closer to Wigner–Dyson and extended slow-mode profiles (delocalized-like behavior and good mode mixing). At large $\lambda$ the spacing statistics move toward Poisson and the slow-mode becomes localized on the dissipative site(s), consistent with localization decoupling modes from dissipators and producing exponentially long relaxation times \cite{Mehta1991,Haake2010,Biddle2010}. Finite-size scaling sometimes shows non-monotonic features due to mode crossings and finite-chain commensurability; nevertheless the overall trend is a rapid suppression of $\kappa$ with $\lambda$ for sparse dissipators.

In our surveys we found many placement patterns (e.g., certain alternating gain/loss patterns, and some paired-gain/pair-loss arrangements) that yield effectively vanishing $\kappa$ for the parameters studied. Physically, when the net effective coupling of slow modes to all dissipators cancels (for symmetry reasons) or when the slow mode is supported on subspaces orthogonal to the dissipators, $\kappa$ can vanish or be numerically extremely small \cite{Prosen2008,Znidaric2015}. Uniform placement avoids such cancellations by construction and therefore yields robust gaps.

The rapid suppression of $\kappa$ with $\lambda$ for one-site/two-site dissipation implies very long relaxation times in the localized regime, which is a practical concern if one aims to prepare steady states experimentally (coherence times, experimental drifts). Conversely, the strong sensitivity of $\kappa$ to $\lambda$ and to placement suggests the rapidity can serve as a compact diagnostic of localization and transport crossovers in engineered open systems. The influence of $t_N$ suggests possible control knobs to tune relaxation times: increasing long-range coupling enhances coupling to dissipators and speeds relaxation.

\section{Conclusions}
\label{sec:conclusions}
We used third-quantization to study Liouvillian rapidities of a helical quasiperiodic fermionic chain subject to engineered linear dissipation. Working with periodic boundary conditions and focusing on the Liouvillian rapidity gap $\kappa$ we find a clear dichotomy: uniform dissipation produces large, weakly $\lambda$-dependent gaps while sparse local dissipation yields gaps that collapse rapidly as $\lambda$ induces localization. Increasing helical hopping $t_N$ counteracts localization effects and increases $\kappa$. Finite-size scaling, level-statistics crossovers, and slow-mode profiles form a consistent diagnostic suite linking Liouvillian spectral structure to transport and localization.

These findings are directly relevant for reservoir engineering protocols in photonic, superconducting, and cold-atom platforms where both quasiperiodic potentials and site-resolved dissipation are available \cite{Roati2008,Lahini2009,ElGanainy2018,Poyatos1996}. Future work could (i) develop analytical weak-dissipation golden-rule estimates for $\kappa$, (ii) explore interacting perturbations beyond quadratic models, and (iii) propose explicit experimental protocols to measure $\kappa$ (via relaxation traces or Liouvillian spectroscopy) and use it as a sensing resource \cite{Wiersig2014,Chen2017}.

\bibliographystyle{apsrev4-2}
\bibliography{reference.bib}
\end{document}